\documentclass[aps,pra,amsmath,amssymb,11pt,final,tightenlines,twoside,twocolumn,nofloats,nofootinbib,superscriptaddress,
showkeys,showkeywords]{revtex4-2}

\usepackage[T2A]{fontenc}
\usepackage[utf8]{inputenc}
\usepackage[russian,english]{babel}
\usepackage{graphicx}

\usepackage{xcolor}

\newcommand{\WI}[2]{#1_{\mathrm{#2}}}

\begin{document}

\selectlanguage{russian}


\keywords{stellar models, differential rotation, polytropes}

\title{Numerical Study of Polytropes with $n = 1$ and Differential Rotation}

\author{\bf \firstname{T.~L.}~\surname{Razinkova}}
\email{razinkova@itep.ru}
\affiliation{National Research Center Kurchatov Institute, Moscow, 123098 Russia}

\author{\bf\firstname{A.~V.}~\surname{Yudin}}
\email{yudin@itep.ru}
\affiliation{National Research Center Kurchatov Institute, Moscow, 123098 Russia}
\affiliation{Novosibirsk State University, Novosibirsk, 630090 Russia}

\author{\bf\firstname{S.~I.}~\surname{Blinnikov}}
\email{blinnikov@bk.ru}
\affiliation{National Research Center Kurchatov Institute, Moscow, 123098 Russia}
\affiliation{Kavli IPMU, Kashiwa, Chiba 277-8583, Japan}

\begin{abstract}
\vspace{3mm}
\vspace{3mm}

The solution space of differentially rotating polytropes with $n=1$ has been studied numerically.
The existence of three different types of configurations: from spheroids to thick tori, hockey puck-like bodies
and spheroids surrounded by a torus, separate from or merging with the central body has been proved. It has
been shown that the last two types appear only at moderate degrees of rotation differentiality, $\sigma\simeq 2$. Rigid-body
or weakly differential rotation, as well as strongly differential, have not led to any ``exotic'' types of configurations.
Many calculated configurations have had extremely large values of parameter $\tau$, which has raised
the question of their stability with respect to fragmentation.
\end{abstract}

\maketitle

\section{INTRODUCTION}
The study of rotating stars, or more generally self-gravitating
bodies, has a long history, apparently dating
back to Newton. The fundamental contributions to
this field by such titans as Poincare and Chandrassekhar
are well known \cite{Tassul}. However, its content
can in no way be considered exhaustive: studies containing
both analytical and numerical approaches to
this problem are still regularly published.

Our study belongs to the second type, and here, we
limited ourselves to considering a purely polytropic
state equation with $n=1$ (the adiabatic index is
$\gamma=2$). It should be noted that rotating polytropes
with $n=1$ attract special attention of theorists because
the equilibrium equations in this case are linear (see
Subsection \ref{sec:EoS} below). In the literature, there are a
sufficient number of (semi-)analytical solutions for this
case (see, for example, \cite{Papoyan1967,Williams1988}) represented by infinite
series in Legendre polynomials both for the case of
rigid-body \cite{James1964} and differential \cite{Blinn1972} rotations. The prefix
``semi'' means that, except in cases of weak rotation,
the coefficients in the analytical expansions have to be
found using one or another numerical method. To
obtain solutions, special techniques, for example, the
introduction of oblate spheroidal coordinates were
also used \cite{Cunningham1977}. There were even studies \cite{Kong2015} whose
authors claimed to have found a \emph{precise} analytical
solution. However, subsequent criticism \cite{Knopik2017} demonstrated
that this solution is only approximate.

A polytrope with $n=1$ was considered as a model
for studying the rotation of Jupiter-type planets and
stars of type $\alpha$--Eridani \cite{Hubbard1975,Kong2013,Vavrukh2020}. We are interested in this
polytrope in the context of continuing research
devoted to neutron stars \cite{Yudin2019}. In this research, we
studied the evolution of low-mass neutron stars with
rigid-body rotation. In this paper, we want to study the
effects of rotation \emph{differentiality} in the most general
form. Although an accurate quantitative study of neutron
stars is obviously only possible within the framework
of the general theory of relativity, certain fundamental
conclusions can be made while remaining
within the framework of a purely Newtonian approximation.
The polytrope with $n=1$ in this context has
another important property (see, for example, \cite{Horedt2004}): as
is known, its radius (in the absence of rotation) is an
eigenvalue of the problem and does not depend on the
mass (or central density). Neutron stars in a wide
range of masses have a similar property \cite{Hensel2007}: their
radius changes very little and is of the order of $13$~km
\cite{Dittmann2024}.

The plan of this paper is as follows: in Section \ref{Base} , the
basic relations are given, the law of rotation is selected
and a numerical method for constructing equilibrium
stellar configurations is described. In Section \ref{slow}, the
case of weakly differential rotation is analyzed. Section
\ref{sigma2} is the main section: it examines the case of $\sigma=2$
in detail (see equation (\ref{rotlaw}) below), and the main
branches of solutions and structures of different types
of calculated configurations are shown. In Section \ref{global},
solutions for different with the aim of looking at the
entire solution space as if ``from above'' and establishing
relationships between different types (branches) of
solutions are provided. In Section \ref{Analytic}, the case $\sigma=2$ is
examined by one of the standard semi-analytical
methods. There, we show that for moderately rotationally
deformed configurations, the results of this
approach coincide with our numerical solutions.
However, it is impossible to describe highly compressed
configurations with them. In conclusion, we
summarize our findings and discuss prospects for further
research.

\section{BASIC DEFINITIONS, PARAMETERS,
AND CALCULATION METHOD}
\label{Base}
We consider an axially--symmetric stellar configuration
in a state of stationary rotation. The equilibrium
equations are written in the form \cite{Tassul,Krat}:
\begin{align}
\frac{1}{\rho}\frac{\partial P}{\partial\xi}&=-\frac{\partial\WI{\varphi}{G}}{\partial\xi}+\omega^2 \xi,\label{dP1}\\
\frac{1}{\rho}\frac{\partial P}{\partial z}&=-\frac{\partial\WI{\varphi}{G}}{\partial z}.\label{dP2}
\end{align}
Here, $P$ is the pressure, $\rho$ is the density of the substance, $\WI{\varphi}{G}$
is the gravitational potential, and $\omega$ is the
angular velocity of rotation. Equation (\ref{dP1}) describes the
equilibrium of matter in a plane perpendicular to the
axis of rotation, and $\xi=\sqrt{x^2+y^2}$ is the cylindrical
radius. In similar equation (\ref{dP2}), $z$ is the coordinate
along the axis of rotation. We consider the case of a
barotropic equation of state, i.e., $P=P(\rho)$. Then,
introducing the enthalpy
\begin{equation}
H(\rho)\equiv\int\limits^{P(\rho)}\frac{dP'}{\rho'},
\end{equation}
and integrating equation (\ref{dP2}), we obtain the Bernoulli
integral:
\begin{equation}
H(\rho)+\WI{\varphi}{G}(\xi,z)=\Omega(\xi), \label{Bernulli}
\end{equation}
where $\Omega(\xi)$ is some function. Substituting this expression
into (\ref{dP1}), we obtain Poincare's theorem on the
constancy of angular velocity on cylindrical surfaces
coaxial to rotation, i.e. $\omega$ must be a function of only
$\xi$, or $\omega=\omega(\xi)$. Hence, then follows the explicit form of
function $\Omega(\xi)$ itself:
\begin{equation}
\Omega(\xi)=\int\limits^{\xi} \omega(\xi')^2\xi' d\xi'+\mathrm{const}.\label{F_xi}
\end{equation}
The equilibrium equation (\ref{Bernulli}) together with expression
(\ref{F_xi}) and the Poisson equation for the gravitational
potential,
\begin{equation}
   \triangle\WI{\varphi}{G} = 4\pi G\rho,\label{Puasson}
\end{equation}
where $\triangle$ is the Laplace operator written for the considered
case in cylindrical coordinates, forms the basis for
solving the problem of a rotating axially symmetric
star.

\subsection{Code \textsf{ROTAT}}
We briefly describe the algorithm implemented in
the \textsf{ROTAT} program \cite{AksBlinn}, which we use to find equilibrium
rotating configurations of stars. It takes Bernoulli
integral (\ref{Bernulli}) as a basis and determines the density,
which is substituted into the right-hand side of
Poisson equation (\ref{Puasson}). The resulting equation contains
only the gravitational potential. This equation is then
written in finite differences on two-dimensional grid
 $(r,\psi)$ , where $r$ is the distance from the center of the
configuration and $\psi$ is the polar angle measured from
the axis of rotation: $\xi=r\sin\psi$ and $z=r\cos\psi$. The
grid is uniform in both coordinates. To calculate the
area outside the sphere that covers the entire star, the
following technique is used \cite{Clement}: a replacement of $r\rightarrow 1/r$
is made and the equations are rewritten using
this transformed variable. This allows calculations to
be carried out in a finite region, and smooth stitching
of solutions is performed on the boundary of the
sphere. At infinity, the condition of $\WI{\varphi}{G} \rightarrow 0$ is set.
Fixing in addition to this the ratio of the polar and
equatorial radii $\theta = \WI{R}{p} / \WI{R}{e}$ and maximal density makes
it possible to determine field $\WI{\varphi}{G}$, and then all other
quantities.

The coefficient matrix for the system of difference
equations is sparse, i.e., it contains many zero elements.
\textsf{ROTAT} uses a method for solving a system of
linear equations (arising from the linearization of the
full system in Newton’s method) with sparse coefficient
matrices that was described in \cite{Esterbu}. The accuracy
of the calculations is controlled by a virial test:
\begin{equation}
VT = \frac{1}{|W|}\Big|2T + W + 3\!\int\! P dV\Big| ,
\end{equation}
where $T$ is the kinetic energy of the star’s rotation, and $W$
is its gravitational energy.

\subsection{Law of Rotation}
First, we define the law of rotation. It is well known
(see, for example, \cite{Tassul}) that function $\omega(\xi)$ cannot be
arbitrary: for rotation stability, the Zolberg–Heyland
condition must be satisfied:
\begin{equation}
\frac{d\omega(\xi)\xi^2}{d\xi}\geq 0,\label{criteriy}
\end{equation}
i.e., the angular momentum per unit mass cannot
decrease during stationary stable rotation. This
inequality motivated us to select from among all the
possibilities the widely accepted law of rotation, which
in our notation is written as follows:
\begin{equation}
\omega(\xi)=\frac{\omega_0}{1+(\sigma{-}1)\frac{\xi^2}{\WI{R}{e}^2}},\label{rotlaw}
\end{equation}
where $\omega_0$ and $\sigma$ are the constants. Parameter $\omega_0$ automatically
determined by the \textsf{ROTAT} code, once configuration
compression $\theta$ is set. Parameter $\sigma$ is set by
the user and determines the degree of differential rotation:
it shows how much faster the center rotates relative
to the last point on the star’s equator at $\xi=\WI{R}{e}$
($\sigma=1$ corresponds, naturally, to solid-body rotation).
Taking into account constraint (\ref{criteriy}), rotation law (\ref{rotlaw})
gives the potentially maximal possible degree of differentiality.

\subsection{Equation of State}
\label{sec:EoS}
As was stated in the Introduction, the state equation
was taken polytrope with $n=1$. The equation of
state is simply $P=K\rho^2$, where $K$ is the constant, and
the enthalpy is then equal to $H(\rho)=2 K\rho$. Linearity of
dependence $H(\rho)$ leads to the fact that Bernoulli
equation (\ref{Bernulli}) also turns out to be a linear relation
between density $\rho$ and gravitational potential $\WI{\varphi}{G}$. Linearity
of the relations between $\rho$ and $\WI{\varphi}{G}$ for $n=1$ follows
from the definition of polytropic index $n$ \cite{Binney2008}.
This fact, along with the linearity of Poisson equation
(\ref{Puasson}), makes it possible, generally speaking, to
exclude $\rho$ (or $\WI{\varphi}{G}$) in the explicit form, and obtain
instead of a system of two equations for two functions
a single equation (the Helmholtz equation) containing
only one function.

The use of a polytropic equation of state also makes
it possible all equations to be made dimensionless by
introducing a characteristic length of $r_0=\sqrt{K/2\pi G}$
and normalizing the density to its maximal value in the
star $\WI{\rho}{m}$. The unit of mass in this case is $M_0=4\pi r_0^3\WI{\rho}{m}$.

As is well known \cite{Horedt2004}, in the absence of rotation
(the spherically symmetric case), for a polytrope with
$n=1$, there is an explicit solution: $\rho=\WI{\rho}{m}\sin(x)/x$.
The radius of a polytrope in dimensionless
coordinates is equal to $\WI{R}{e}=\pi$. It is easy to show that
the dimensionless mass $\WI{M}{s}$ is also
equal $\pi$. Below, we will use exactly these dimensionless
quantities everywhere.

\section{WEAKLY DIFFERENTIAL ROTATION}
\label{slow}
We begin the description of the obtained results
with the case of relatively small rotation differential
parameter $\sigma$.
\begin{figure*}[htb]
\includegraphics[width=1.\textwidth]{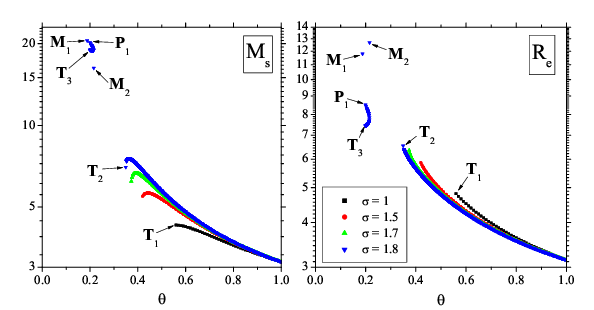}
\caption{Mass of a star $\WI{M}{s}$ (left) and its equatorial radius $\WI{R}{e}$ (right) as functions of parameter $\theta$ for several $\sigma$ values. The arrows
indicate the configurations whose structures are shown in the graphs below.}
\label{Pix:MRlow}
\end{figure*}
Figure \ref{Pix:MRlow} shows the calculated dependences of
dimensionless masses $\WI{M}{s}(\theta)$ (left) and radius $\WI{R}{e}(\theta)$
(right) for four $\sigma$ values: from $\sigma=1$ (solid body rotation)
to $\sigma=1.8$. Configurations related to different $\sigma$
differ in both color and type of symbols. All series of
configurations start, obviously, from the point of $\WI{M}{s}=\WI{R}{e}=\pi$
at $\theta=1$ (no rotation) and reach as $\sigma$
grows to ever smaller $\theta$ values, where they break off.

\begin{figure}
\includegraphics[width=0.5\textwidth]{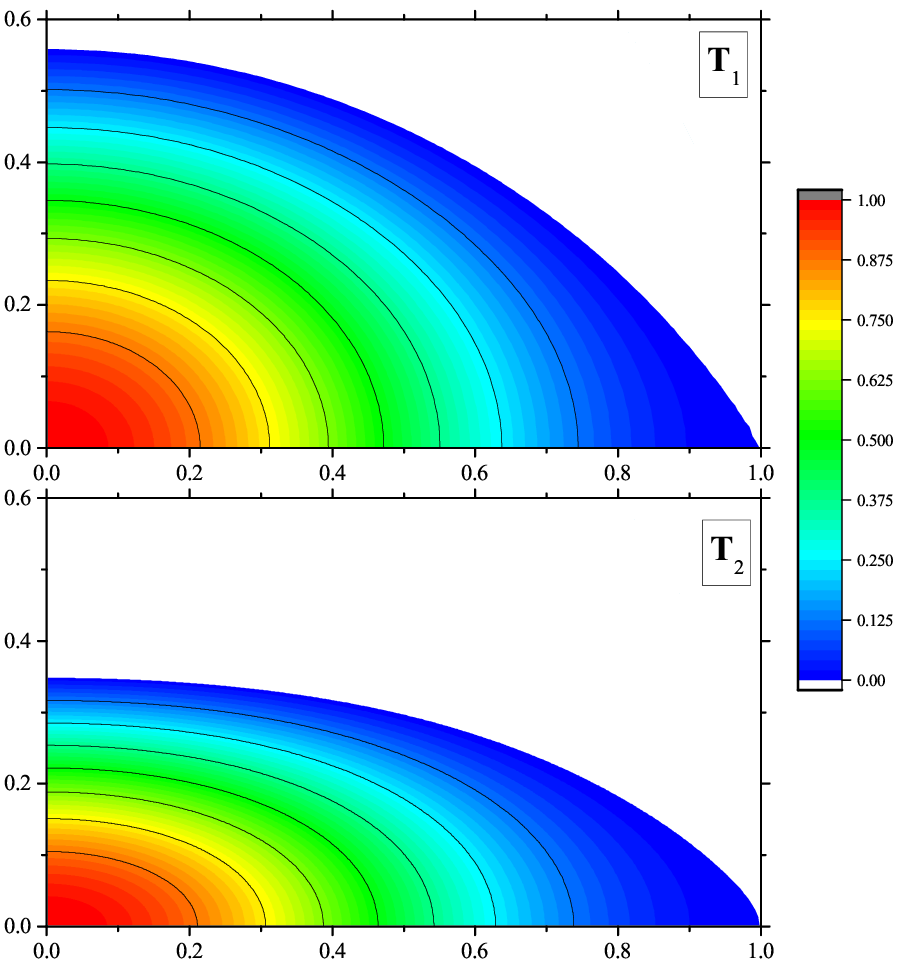}
\caption{Structure of configurations of $\mathbf{T}_1$ ($\sigma=1$) and $\mathbf{T}_2$ ($\sigma=1.8$) marked by arrows in Fig.~\ref{Pix:MRlow} in normalized coordinates $\{\xi/\WI{R}{e},z/\WI{R}{e}\}$. Details are in the text.}
\label{Pix:LowT}
\end{figure}

Examples of the fastest rotating stars in these \emph{continuous}
series for the case of $\sigma=1$ (configuration $\mathbf{T}_1$)
and the case of $\sigma=1.8$ ($\mathbf{T}_{2}$) are shown in Fig.~\ref{Pix:LowT}. Here,
the lines of the density level normalized to the maximum (see the color scale on the right) in coordinates
in which the equatorial radius of the star is also normalized
to unity, i.e., in coordinates $\{\xi/\WI{R}{e},z/\WI{R}{e}\}$, are
shown. Only a quarter of the entire configuration is
shown: the star is symmetrical both relative to the
abscissa axis (i.e., relative to the equatorial plane due
to Lichtenstein’s theorem, see, for example, \cite{Tassul}) and
relative to the ordinate axis (the axis of rotation).

We return to Fig.~\ref{Pix:MRlow}. Obviously, as $\sigma$ grows, the
maximal flattening of the configurations increases,
and this is exactly what should be expected. However,
at $\sigma=1.8$, something unexpected happens: after the
gap of $0.216\leq\theta\leq 0.35$, in which there are no solutions, two new types of configurations suddenly
appear: \textbf{P} and \textbf{M}-type. They lie within the same narrow
range of $\theta$ values, and here, the ambiguity (degeneracy)
of the solution for this parameter is manifested.

\begin{figure}[htb]
\includegraphics[width=0.5\textwidth]{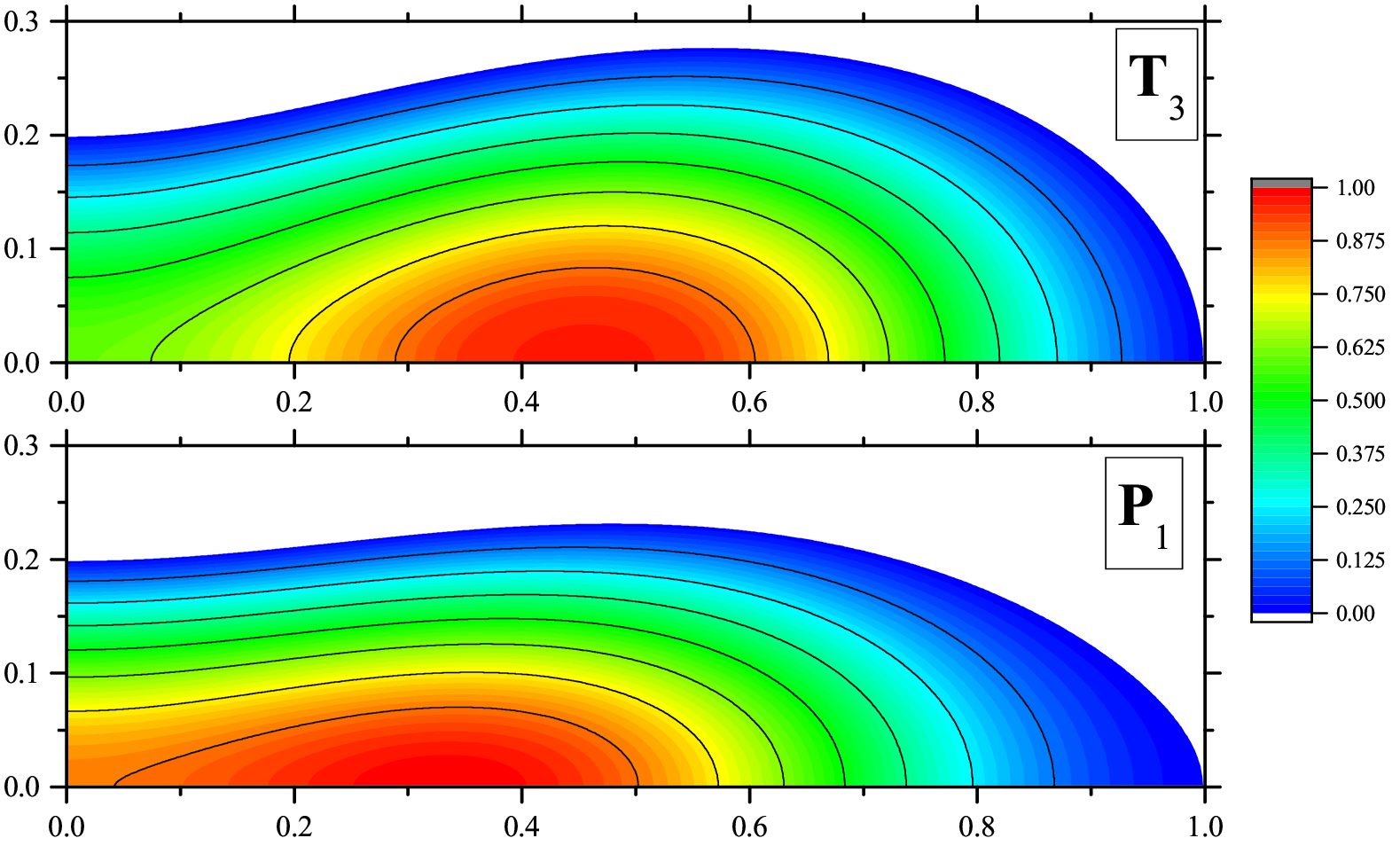}
\caption{Structure of configurations $\mathbf{T}_3$ and $\mathbf{P}_1$ (see Fig.~\ref{Pix:MRlow}).}\label{Pix:LowP}
\end{figure}

We first consider a series of configurations that
connect $\mathbf{T}_3$ and $\mathbf{P}_1$. Its marginal structures are shown in
Fig.~\ref{Pix:LowP}.
It is evident that, despite the almost identical flattening
$\theta\approx 0.2$, configuration $\mathbf{T}_3$ is closer to the torus
(hence, the name of this type), and $\mathbf{P}_1$ is more uniformly
compressed and flat, resembling a hockey
puck. It is interesting to note that these configurations
(especially $\mathbf{T}_3$) resemble the ``concave hamburger'' obtained in \cite{Hachisu1982}. We will see later that, generally
speaking, configurations $\mathbf{T}$ and $\mathbf{P}$ are indeed separate
and easily distinguishable types, but here, we observe
their smooth transition into each other. The reason for
this is discussed below in Section \ref{global}.

\begin{figure}
\includegraphics[width=0.5\textwidth]{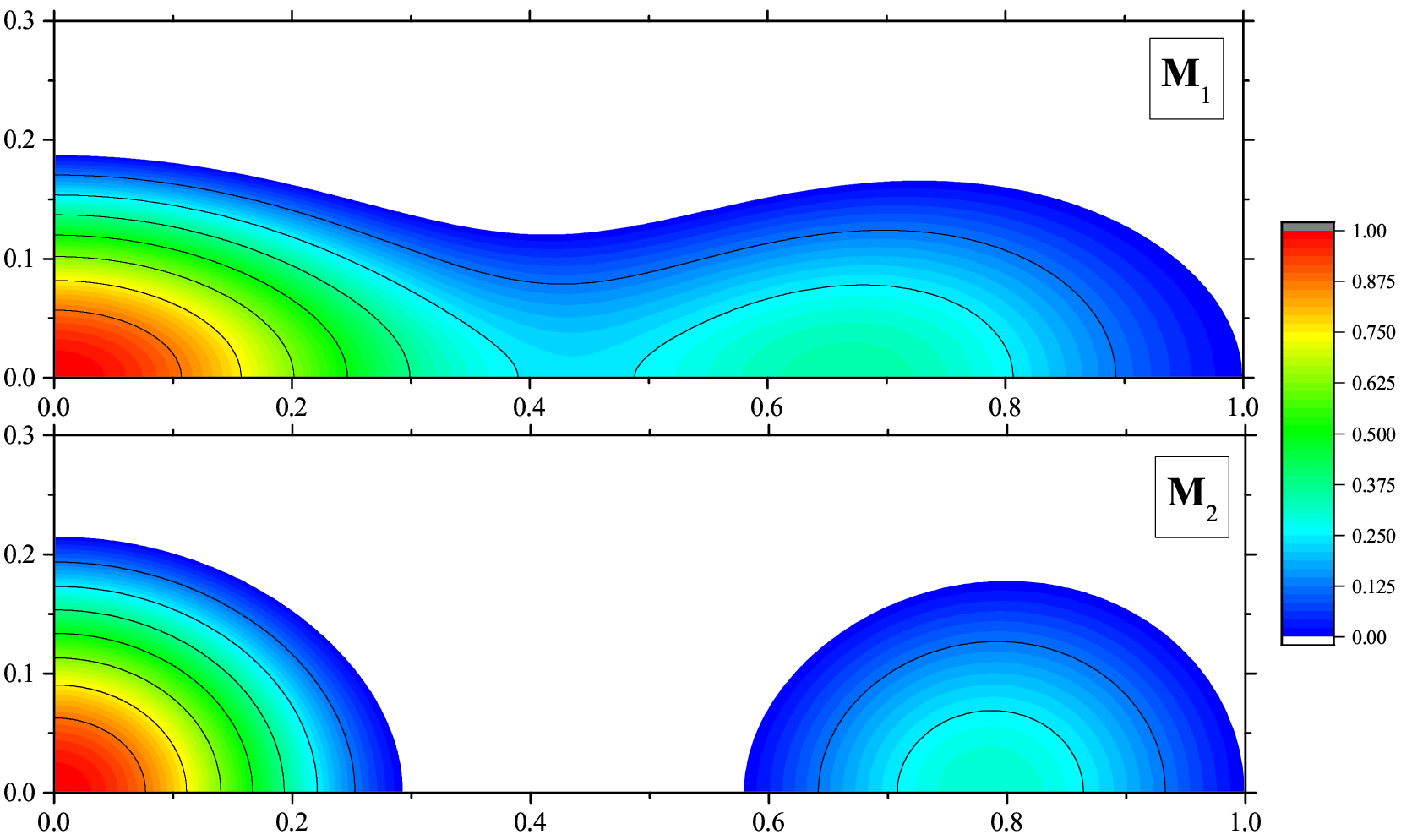}
\caption{Structure of configurations $\mathbf{M}_{1,2}$ (see Fig.~\ref{Pix:MRlow}).}
\label{Pix:LowM}
\end{figure}

Finally, we move on to an even more surprising
type of configuration --- type $\mathbf{M}$. Unlike types $\mathbf{T}$ and $\mathbf{P}$,
here, we do not even have a series, but only two solutions!
Their structures are shown in Fig.~\ref{Pix:LowM}.
$\mathbf{M}_1$ is a structure similar to a matryoshka (russian doll) lying
on its side (hence, the ``M'' is in the name of this type),
and $\mathbf{M}_2$ is a central body surrounded by a torus. It is
interesting to note that configuration $\mathbf{M}_2$ is extremely
similar to the spheroid-ring systems studied in \cite{Basillais2019},
although for the case of solid-body rotation and
incompressible fluid.

It is obvious that between $\mathbf{M}_1$ and $\mathbf{M}_2$, it is possible
to imagine a continuous transition, but despite all our
tricks, we were unable to detect intermediate configurations.
This may be due to the above-mentioned
degeneration of solutions with respect to parameter $\theta$:
all intermediate solutions are ``occupied'' with a series
from $\mathbf{T}_3$ to $\mathbf{P}_1$.

\section{CASE $\sigma=2$}
\label{sigma2}
\begin{figure*}
\includegraphics[width=1.\textwidth]{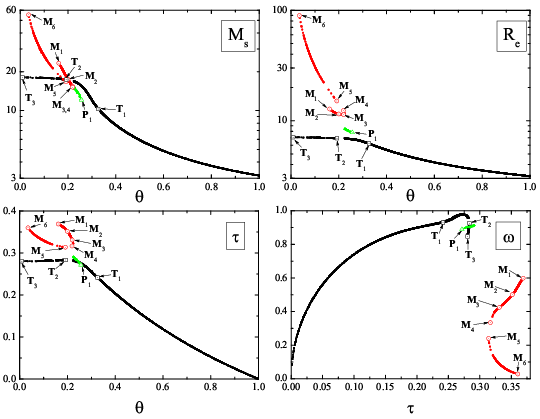}
\caption{Dependencies $\WI{M}{s}(\theta)$, $\WI{R}{e}(\theta)$, $\tau(\theta)$, and $\omega(\tau)$. Configurations of different types are highlighted in color. The large symbols
with markings show the configurations shown below in Figs.~\ref{Pix:Tor}, \ref{Pix:Puck}, and \ref{Pix:Matryoshka}. Other details are in the text.}\label{Pix:MAIN2}
\end{figure*}

We consider the option with $\sigma=2$, i.e., the case
when the center of the star rotates twice as fast as the
last point of the star on the equator at $r=\WI{R}{e}$. This case
is the most interesting since it not only contains all the
types of rotating configurations we found, but is also
characterized by a very non-trivial topology of curves
in the space of the considered parameters. The latter
are shown in Fig.~\ref{Pix:MAIN2}.

The four panels of this figure show the following
dependencies.

Top left: mass $\WI{M}{s}$ as a function of the configuration
compression parameter $\theta=\WI{R}{p}/\WI{R}{e}$.

Top right: equatorial radius $\WI{R}{e}(\theta)$.

Bottom left: $\tau(\theta)$, where $\tau\equiv\WI{E}{rot}/|\WI{W}{g}|$ is the
dimensionless parameter equal to the ratio of the
kinetic energy of rotation of a star $\WI{E}{rot}$ to the modulus
of its gravitational energy $\WI{W}{g}$. Quantity $\tau$ determines
the importance of rotation effects and, in particular, is
a critical indicator in the question of fragmentation of
a rapidly rotating stellar configuration \cite{Tassul}.

Bottom right: $\omega(\tau)$, where $\omega=J/I$ is the ``average''
angular velocity of rotation that is defined as the
ratio of total angular momentum $J$ of configuration to its
inertia momentum $I$. For a rigid body rotating star, $\omega$
coincides with the angular velocity of rotation. Let us
recall that all quantities considered here are dimensionless,
in particular, is measured in units of $\sqrt{G\WI{\rho}{m}}$.

The colors indicate configurations that belong to
different types. We begin our consideration with configurations
of type \textbf{T} shown in black. They start with
non-rotating configurations, for which $\WI{R}{e}=\WI{M}{s}=\pi$.
At the same time, obviously, $\tau=\omega=0$. As compression increases ($\theta$ decreases), all the considered parameters
initially increase monotonically.

This continues until the value of $\theta\approx 0.25$, which is
a bifurcation point: other types of configurations
appear here, \textbf{P} and \textbf{M}, which we will describe later. It
is interesting to note that the value of parameter $\tau$ is
close in this case to the critical value $\tau\approx 0.27$ of the
onset of dynamic instability with respect to fragmentation
(see, for example, \cite{Tassul}). After that, growth of $\WI{M}{s}$, $\WI{R}{e}$
and $\tau$ on the branch of \textbf{T}-type slows down significantly
or stops completely, and parameter $\omega$, reaching
its maximum with $\omega\approx 1$, begins to decrease.

\begin{figure}
\includegraphics[width=0.5\textwidth]{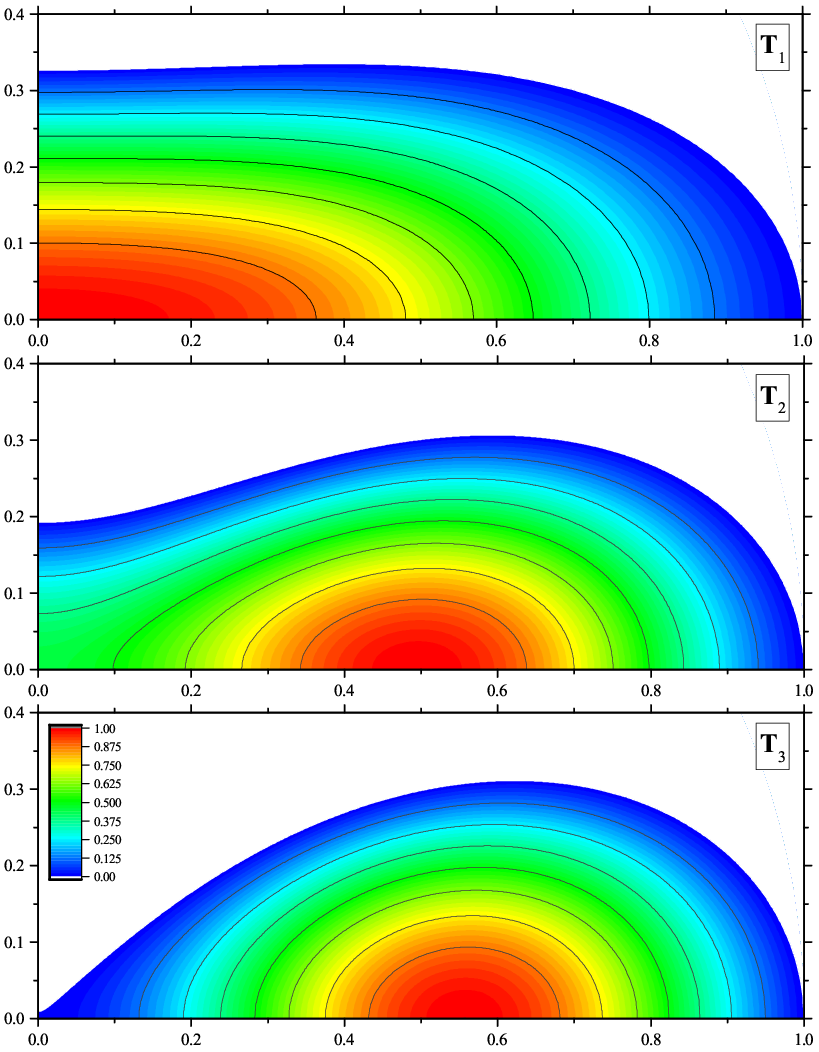}
\caption{Density level lines in configurations $\mathbf{T}_1{-}\mathbf{T}_3$ (see Fig.~\ref{Pix:MAIN2})}\label{Pix:Tor}
\end{figure}

\begin{figure}
\includegraphics[width=0.5\textwidth]{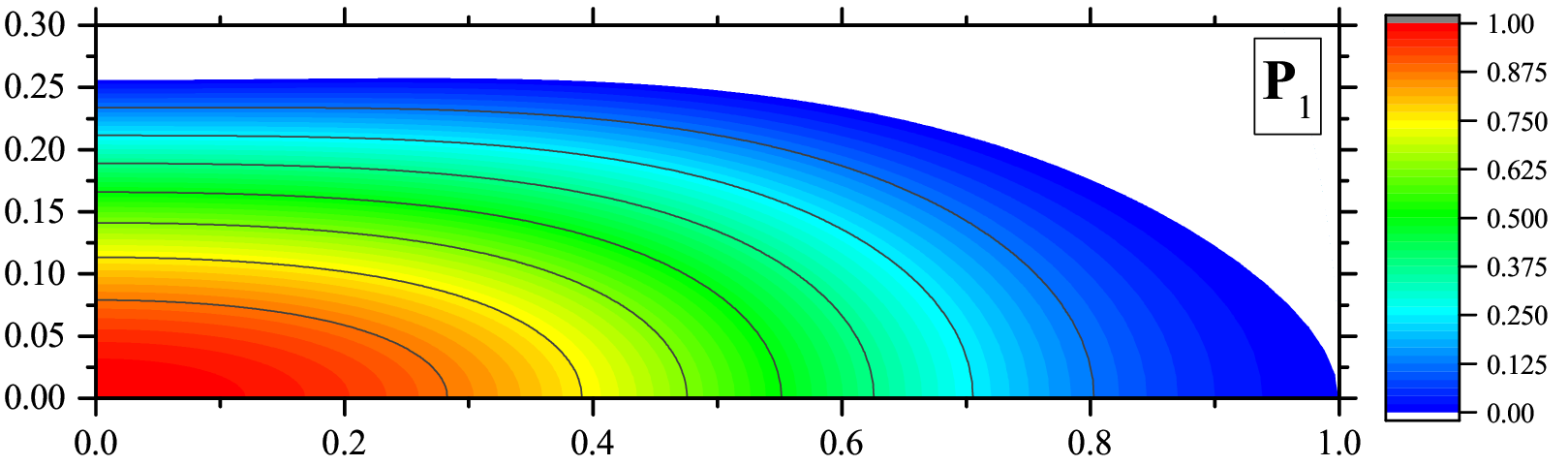}
\caption{Density level lines in configuration $\mathbf{P}_1$ (see Fig.~\ref{Pix:MAIN2}).}\label{Pix:Puck}
\end{figure}

What are configurations of type \textbf{T} and how do they
change with the considered decrease in parameter $\theta$?
The answer is given in Fig.~\ref{Pix:Tor}.
It shows the density level lines for configurations
from $\mathbf{T}_1$ to $\mathbf{T}_3$ (see Fig.~\ref{Pix:MAIN2}). Obviously, non-rotating
configurations at $\theta=1$ are spherically symmetric.
Then, as the rotation rate increases, the configurations
become increasingly flattened at the poles. Configuration $\mathbf{T}_1$
shows a structure that is already sufficiently
compressed and flat. It is also interesting to compare it
with the structure of type \textbf{P} (see Fig.~\ref{Pix:Puck} and its
discussion).

When moving towards $\mathbf{T}_2$, an important phenomenon
occurs: the place, where the density reaches its
maximal value, shifts from the center of the star to the
periphery. The central region remains sufficiently
``thick''. Ultimate configuration $\mathbf{T}_3$ corresponding to
the value of $\theta=0$ is a torus. It is important to note the
difference between this torus and those toroidal structures
described in the literature (see, for example, \cite{Ostriker1964,Wong1974}): it is thick, i.e., the main radius $R$ of the torus
comparable to its minor radius $r$. It is worth mentioning
here that Wong worked with an incompressible
fluid, while Ostriker, although considered, in particular,
polytropes, used approximation $R\gg r$ everywhere.

We now consider configurations of other types that
appear in calculations after bifurcation. We start with
structures of type \textbf{P}. They are shown in green in Fig.~\ref{Pix:MAIN2}.
As can be seen, they occupy a very small area of
parameters in all diagrams, which is why in Fig.~\ref{Pix:Puck} we
show the structure of only one of them (they all are
very similar). These structures resemble a $\mathbf{T}_1$ flattened
one and a half times (see Fig.~\ref{Pix:Tor}). Despite this similarity,
an examination of the various integral parameters
shown in Fig.~\ref{Pix:MAIN2} shows that configurations of
type \textbf{P} represent a \emph{separate} branch with its own unique
properties.

\begin{figure*}
\includegraphics[width=1.\textwidth]{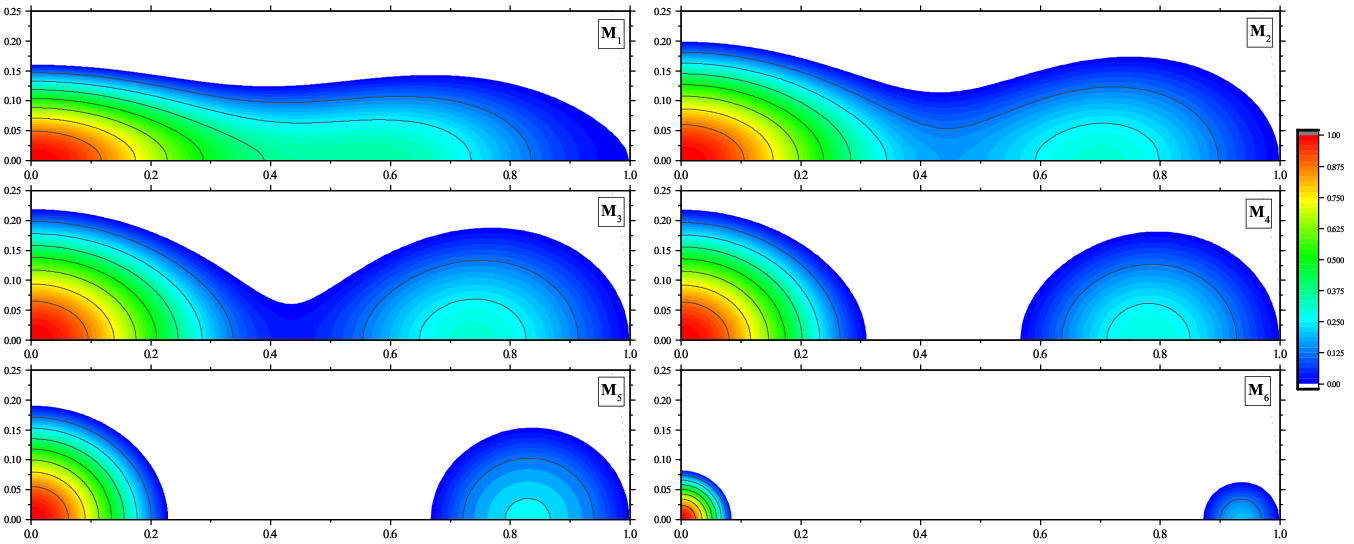}
\caption{Density level lines in configurations  $\mathbf{M}_1{-}\mathbf{M}_{6}$(see Fig.~\ref{Pix:MAIN2}).}\label{Pix:Matryoshka}
\end{figure*}

We finally move on to the description of the
extremely interesting branch with the \textbf{M} marker
(matryoshka doll). The reason for this, the name
becomes clear from studying Fig.~\ref{Pix:Matryoshka}.
We start with configuration $\mathbf{M}_1$. It is something that
vaguely resembles structures $\mathbf{T}_1$ or $\mathbf{P}_1$, but with a more
pronounced core and periphery separated by a noticeable
deflection. Moving on through $\mathbf{M}_2$ to $\mathbf{M}_3$, this
deflection increases, and the section connecting the
core with the periphery finally breaks during the transition
to $\mathbf{M}_4$. All subsequent configurations, including
$\mathbf{M}_4$, represent a central body surrounded by a torus.
Moreover, when moving towards $\mathbf{M}_6$ the rotation
becomes slower and slower, which is evident from the
central part approaching spherical symmetry and the
drop in parameter $\omega$. The radius of the torus increases,
making such configurations already significantly similar
to those considered by Ostriker~\cite{Ostriker1964}.

Despite its low density, this torus contains most of
the mass of the system and significant angular
momentum, as well as rotational energy, which is evident
from the growth of parameter $\tau$. It is important to
note that \emph{all} \textbf{M}-configurations have extremely high
values of $\tau>0.3$, approaching in configurations $\mathbf{M}_1$
and $\mathbf{M}_6$ to the values of $\tau\simeq 0.36$. According to \cite{Bardeen1971,BodenOstr1973}, at this value, the rotating configuration becomes
secularly unstable with respect to axially symmetric
disturbances. It seems extremely interesting to study
the stability of these structures, which, obviously, can
only be done by performing a dynamic 3D-calculation.
The authors are currently working on its implementation.

Again, it is worth emphasizing that the structure of
solutions turns out to be degenerate with respect to
parameter $\theta$: as can be seen in Fig.~\ref{Pix:MAIN2}, different solutions
may correspond to one $\theta$ value. On the other
hand, it is noteworthy that in the range of values of
another important parameter of $0.29\lesssim\tau\lesssim 0.31$,
there are no solutions at all.

\section{GLOBAL DESCRIPTION}
\label{global}
\begin{figure*}
\includegraphics[width=1.\textwidth]{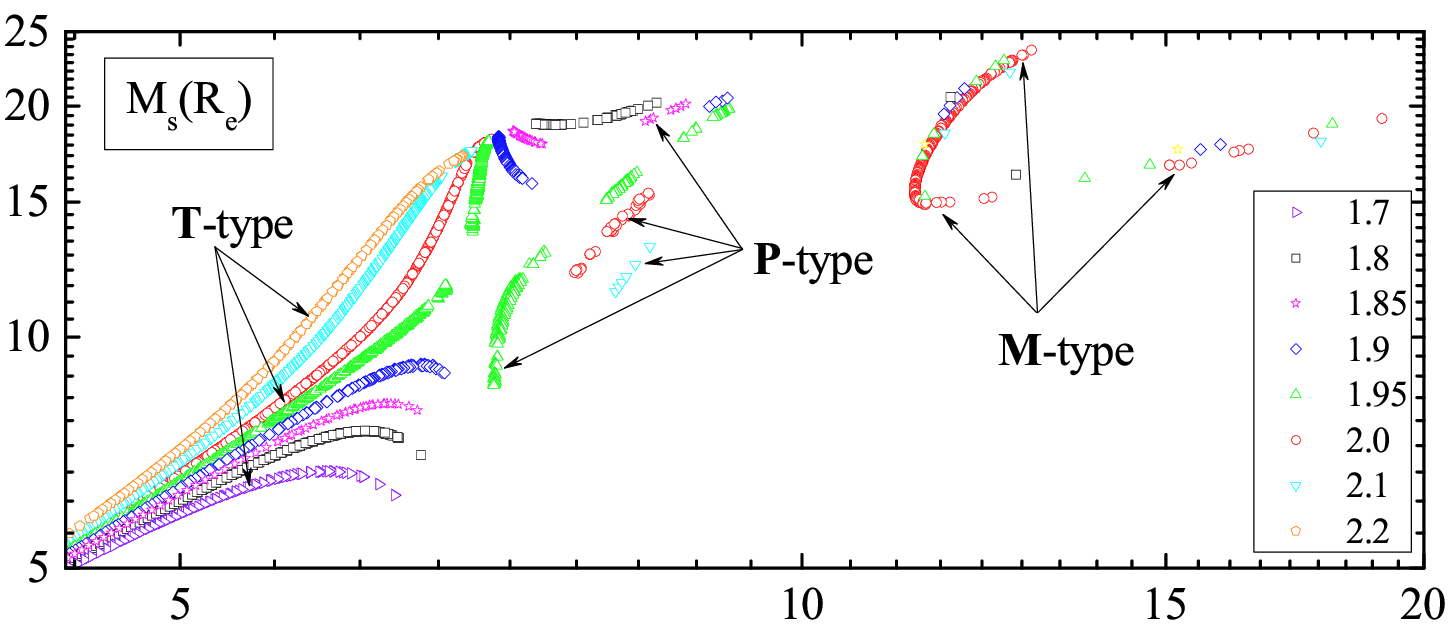}
\caption{Diagram $\WI{M}{s}-\WI{R}{e}$ for several values of parameter $\sigma$.}\label{Pix:ViewFromTop}
\end{figure*}

Despite the above-discussed case of relatively weak
rotation and the special and most interesting case of
$\sigma=2$, the question about what happens globally with
a gradual increase in parameter $\sigma$ characterizing the
degree of differential rotation still remains. What is
needed here is, so to speak, a ``top view'' that would
help to distract oneself from the particular characteristic
of each specific value of $\sigma$ and see the whole picture.
This purpose, Fig.~\ref{Pix:ViewFromTop} shows the ``Mass–Radius''
diagram for several $\sigma$ values.

We start with the area of small $\sigma$. In the lower left
corner of the figure, the purple triangles show the case
of $\sigma=1.7$, which is also shown in Fig.~\ref{Pix:MRlow}. There are no
``exotic'' configurations here. They appear at $\sigma=1.8$
(black squares) as a part of a ``parabola'' with branches
upwards at $\WI{M}{s}\approx 20$ and $\WI{R}{e}\sim 7$ (it corresponds to
configurations between $\mathbf{T}_3$ and $\mathbf{P}_1$ in Fig.~\ref{Pix:MRlow}) and in the
form of two solutions of type \textbf{M} falling on the branch
of ``matryoshka'' on the right ($\WI{M}{s}\gtrsim 15, \WI{R}{e}\gtrsim 11$).
Already at $\sigma=1.85$ (purple stars), the upper ``parabola''
splits into two branches. Thus, a small ``parabola''
at $\sigma=1.8$ is the only bridge connecting types \textbf{T}
and \textbf{P}.

It is interesting to follow the evolution of these
branches as $\sigma$ grows. The left branch of the parabola
becomes more and more vertical and at $\sigma\sim 1.95\div 2$
merges with the lower branch coming from non-rotating
configurations, thus forming configurations of
type \textbf{T}. The right branch obviously goes into configurations
of type \textbf{P}, reaching its greatest prevalence at $\sigma=1.95$
and gradually fading away to $\sigma=2.1$.

Separately, it is necessary to consider the branch of
configurations of type \textbf{M}. It is most fully represented
by the case of $\sigma=2$. It is worth noting that the right
side of the branch of \textbf{M}-configurations extend far to
the right and up beyond the area of the figure and is
not shown simply to save space: such configurations
represent a slowly rotating central body surrounded by
a torus that is increasingly distant from it (see configuration $\mathbf{M}_6$
in Fig.~\ref{Pix:Matryoshka}). It is interesting that \textbf{M} -configurations
for all $\sigma$ lie practically on the same branch,
although there is no point in considering completeness
or continuity in the sequence of configurations with
different $\sigma$: taken separately, \textbf{M}-solutions for $\sigma\neq 2$
represent, at first glance, a disparate set of points.

It should also be noted that at further growth of $\sigma$,
no other ``exotic'' configurations arise, and the entire
sequence of solutions for this $\sigma$ represents one branch
leading from non-rotating to \textbf{T}-configurations.

It is appropriate here to discuss the completeness of
the presented solutions. In many branches in Fig.~\ref{Pix:ViewFromTop},
breaks are visible, and \textbf{M}-solutions, as was already
mentioned above, are often presented only as individual,
single configurations. In our opinion, this seems
to be a consequence of the already mentioned degeneracy
of solutions with respect to input parameters, in
particular, with respect to $\theta$. The Newton iteration
algorithm implemented in the \textsf{ROTAT} code in this
case begins to ``jump'' between solutions, demonstrating
poor convergence or lack thereof. Therefore, we
cannot be absolutely sure of the completeness of the
presented solutions. However, we can be sure of their
\emph{reality}. In fact, we managed using the well-known
astrophysical code \textsf{FLASH} \cite{FLASH} to carry out 3D modeling
of some of the above-considered ``exotic'' configurations
of different types with extreme values of
parameters, such as $\theta$ and $\tau$. And all of them demonstrate
hydrodynamic stability for at least tens of revolutions.
This indicates that the hydrostatic equations
for the initial configurations were solved correctly by
the {ROTAT} code. The authors plan to publish the
results of this study in the near future.

\section{ANALYTICAL APPROACH}
\label{Analytic}
In light of some unusualness of some of the above
results, their independent verification seems necessary.
Below, we show the results of a standard analytical
approach dating back to Chandrasekhar \cite{Chandra1962}.

After the transition to dimensionless variables, the
system of equations consisting of equilibrium equation
(\ref{Bernulli}) and Poisson equation (\ref{Puasson}) looks as follows:
\begin{align}
&\rho+\varphi=\Omega(\xi),\label{rot_eq_rho}\\
&\triangle\varphi=\rho.\label{rot_eq_puass}
\end{align}
Due to the already noted linearity of the equations,
this system is equivalent to one Helmholtz-type equation
with respect to one of variables: $\rho$ or $\varphi$.

\begin{figure}
	\begin{center}
		\includegraphics[width=0.45\textwidth]{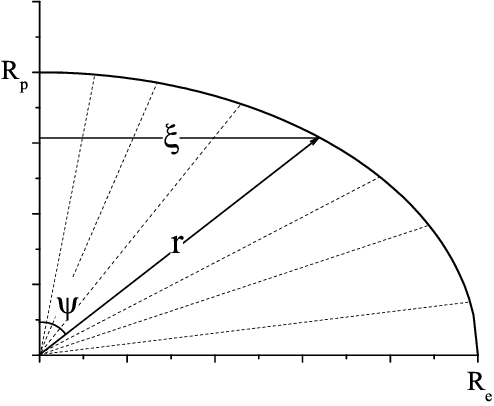}
	\end{center}
	\vspace{-1cm}
	\caption{Schematic representation of the computational domain and the boundary of the star.} \label{Fig.illustrative}
\end{figure}
In the case of relatively weak rotation, when the
configuration of the star does not differ much from
spherical, it seems natural to look for a solution in $(r,\mu)$
coordinates (where $\mu=\cos\psi$, see Fig.~\ref{Fig.illustrative}) in the
form of an expansion in Legendre polynomials $P_k(\mu)$.
The Laplace operator in this case is:
\begin{equation}
\triangle\varphi=\frac{1}{r^2}\frac{\partial}{\partial r}\!\!\left(\! r^2\frac{\partial\varphi}{\partial r}\!\right)+
\frac{1}{r^2}\frac{\partial}{\partial\mu}\!\!\left(\!(1{-}\mu^2)\frac{\partial\varphi}{\partial\mu}\!\right).\label{Laplace_operator}
\end{equation}
Arbitrary function $f=f(r,\mu)$ can be expanded in
a row:
\begin{equation}
f(r,\mu)=\sum\limits_{k=0}^{\infty}f_{2k}(r)P_{2k}(\mu),\label{Legendre_series}
\end{equation}
where it is taken into account that, due to the symmetry
of the problem, the sum in (\ref{Legendre_series}) can contain only
even indices. The orthogonality properties of Legendre
polynomials make it possible to find explicitly
functions $f_{2k}(r)$:
\begin{equation}
f_{2k}(r)=\frac{4k{+}1}{2}\!\int\limits_{-1}^1f(r,\mu)P_{2k}(\mu)d\mu
\end{equation}
If we now express $\rho$ from equation (\ref{rot_eq_rho}) and substitute
it into (\ref{rot_eq_puass}), then, taking into account the properties of
Legendre polynomials, we obtain the following equations
for the components of the gravitational potential:
\begin{equation}
\varphi''_{2k}+\frac{2}{r}\varphi'_{2k}+\left[1{-}\frac{2k(2k{+}1)}{r^2}\right]\varphi_{2k}=\Omega_{2k}(r),\label{Bessel_spherical}
\end{equation}
where the primes denote derivatives with respect to $r$.
The solution to the homogeneous equation is spherical
Bessel functions $j_{2k}(r)$ and $y_{2k}(r)$, and, due to the
limitations of $\varphi$ at zero, only the first ones are suitable
for us. We seek the solution of the inhomogeneous
equation by the method of variation of constants, i.e.,
in the form of $\varphi_{2k}=c_{2k}(r)j_{2k}(r)$. Then, for $c_{2k}$, we
obtain the equation:
\begin{equation}
\left(r^2j^2_{2k}c'\right)'=r^2\Omega_{2k}j_{2k}.\label{c_spherical}
\end{equation}
The solution to this equation is a double integral:
\begin{equation}
c_{2k}=\int\limits^r\!\frac{dr_1}{r_1^2j_{2k}^2(r_1)}\!\int\limits^{r_1}\!r_2^2\Omega_{2k}(r_2)j_{2k}(r_2)dr_2.\label{c_2k_2int}
\end{equation}
To simplify it, we use the following property of the
Wronskian of the spherical Bessel functions \cite{NIST}:
\begin{equation}
W\big(j_k(r),y_k(r)\big)=
\begin{vmatrix}
j_k & y_k\\
j'_k & y'_k
\end{vmatrix}
=\frac{1}{r^2}.
\end{equation}
With its help, expression (\ref{c_2k_2int}) can be reduced to a single
integral, and the sought components of gravitational
potential will $\varphi_{2k}=c_{2k}j_{2k}$ be written in the
form:
\begin{multline}
\varphi_{2k}(r)=
\int\limits_0^r\! r_1^2\Omega_{2k}(r_1)\big[y_{2k}(r)j_{2k}(r_1)-\\
-y_{2k}(r_1)j_{2k}(r)\big]dr_1+C_{2k}j_{2k}(r),\label{phi_2k}
\end{multline}
where $C_{2k}$ are some constants. From the condition of
normalization at zero: $\rho(0)=1$ and the fact that $\Omega(0)=0$, we obtain that $C_0=-1$.

\subsection{External Solution and Stitching of Solutions}
It is well known that the solution of the Poisson
equation in a vacuum, regular at infinity, has an
explicit solution for components (\ref{Legendre_series}) of the potential:
\begin{equation}
\varphi_{2k}(r)=A_0\delta_{k0}+\frac{B_{2k}}{r^{2k{+}1}},\label{phi_out}
\end{equation}
where $\delta_{k0}$ is the Kronecker symbol, and $B_{2k}$ are the
constants.

At the edge of the star, density $\rho$ turns to zero.
Besides, potential $\varphi$ and its gradient are continuous.
The problem is that the shape of this boundary is not
known in advance. We consider a method for numerically
solving this problem (see, for example, \cite{Papoyan1967}).

Let the angular velocity of rotation $\omega(\xi)$, and therefore,
potential $\Omega$ be known. We restrict ourselves in
expansions (\ref{Legendre_series}) to the $N$ order, i.e., $k=(0\div N)$. For
internal potential (\ref{c_2k_2int}), hereinafter designated by the
index $\mathrm{in}$, there is $N$ unknown coefficients $C_{2k}$
because $C_0=-1$ is specified. For the external (index
(индекс $\mathrm{out}$)) number of unknowns is $N+2$ ($A_0$ and $N+1$ of
coefficients $B_{2k}$). We select $N+1$ directions set by
angles $\mu_i$, as is shown in Fig. (\ref{Fig.illustrative}) by dashed lines. Then,
we will have $N+1$ unknown coordinates $r_i$ of the star’s
surface corresponding to their angle $\mu_i$. In total, we
obtain the following $3(N+1)$ unknowns. Each of $N+1$
directions produces three equations:
\begin{align}
&\varphi^{\mathrm{in}}(r_i,\mu_i)=\sum\limits_{k=0}^N\varphi^{\mathrm{in}}_{2k}(r_i)P_{2k}(\mu_i)=\Omega(\xi_i),\label{phi_in_shiv}\\
&\varphi^{\mathrm{out}}(r_i,\mu_i)=\sum\limits_{k=0}^N\varphi^{\mathrm{out}}_{2k}(r_i)P_{2k}(\mu_i)=\Omega(\xi_i),\label{phi_out_shiv}\\
&\frac{d}{dr}\varphi^{\mathrm{in}}(r,\mu_i)\Big|_{r=r_i}=\frac{d}{dr}\varphi^{\mathrm{out}}(r,\mu_i)\Big|_{r=r_i}\label{phi_grag_shiv},
\end{align}
where, obviously, $\xi_i=r_i\sqrt{1{-}\mu_i^2}$.

To calculate the derivative of quantities $\varphi^{\mathrm{in}}_{2k}$, the
properties of Bessel functions should be used: $f'_n=f_{n{-}1}-(n{+}1)f_n/x$
and $f'_0=-f_1$, where $f=j$
or $f=y$.

\subsection{Results of This Approach}
\begin{figure*}
	\begin{center}
		\includegraphics[width=1.\textwidth]{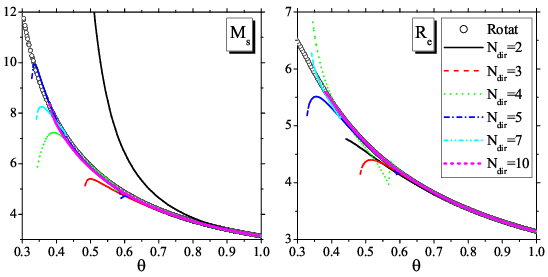}
	\end{center}
	\vspace{-1cm}
	\caption{Comparison of results of numerical approaches for $\sigma=2$. On the left is the mass $\WI{M}{s}$ of the star, and on the right is the
equatorial radius $\WI{R}{e}$. Other details are in the text.} \label{Fig.calc_2}
\end{figure*}

We illustrate the results of the numerical approach
described above using the example of the case of $\sigma=2$. Figure \ref{Fig.calc_2} shows mass $\WI{M}{s}$ of a star (left) and its
equatorial radius $\WI{R}{e}$ (right) as functions of $\theta$.

Empty circles show calculations using the \textsf{ROTAT}
program. Lines of different colors and types show calculations
according to the algorithm described above
for several values of $\WI{N}{dir}$, numbers of basic directions,
namely, $\WI{N}{dir}=2,3,4,5,7,10$. For each $\WI{N}{dir}$, the
angle distribution of directions is uniform.

The simplest case of $\WI{N}{dir}=2$ (namely, the directions
to the pole and the equator), as can be seen, is
accurate only up to $\theta\gtrsim 0.85$. At lower $\theta$, it systematically
and greatly overstates the value of mass,
which is simply taken $B_0$ with the opposite sign (see
formula \ref{phi_out}). Other values $\WI{N}{dir}$ reproduce the parameters
of stars much more accurately, and, obviously,
the greater $\WI{N}{dir}$, the better. However, they all face a
certain peculiarity: when the value $\theta$ little less than
$0.6$, a numerical instability arises, which either leads to
a ``jump'' to another solution (see, for example, the
curves for $\WI{N}{dir}=4,5$) or to significantly worse convergence
of iterations. This property is also characteristic
of the calculations we performed for other $\sigma$. In
general, it is not surprising that the smaller the value $\theta$
we take, the worse the convergence, and the greater
the number $\WI{N}{dir}$ must be taken to obtain correct
results.

Using this method, we were unable to reproduce
highly compressed configurations with $\theta\lesssim 0.4$ (the
so-called ``exotic''). This is not surprising: the lack of
convergence of expansions in spherical harmonics in
the case of strongly deformed configurations is widely
known (see, for example, \cite{Balmino1994,Bevis2024}). To analyze them,
algorithms of a different type are needed.

\section{CONCLUSIONS}
In this paper, we have undertaken a detailed
numerical study of polytropes with $n=1$ and differential
rotation. It should be noted here that conventional
(semi-)analytical methods are usually based on expansions
in some small parameters and are not applicable
to the case, where the rotation is strong. We have
obtained configurations of various types, with different topologies, and often with extreme values of such
parameters as compression $\theta$ or fragmentation parameter$\tau$.

We have identified the existence of three different
types of configurations. Firstly, it is type \textbf{T}: a branch of
solutions from non-rotating, spherically symmetric
stars, through increasingly flattened objects to thick
tori with $\theta\simeq 0$. Secondly, it is a small branch of
type \textbf{P}: puck-like configurations distantly related to
type \textbf{T}. This becomes clear only when studying configurations
with $\sigma\sim 1.8$ (see the topology of the solution
space in $\WI{M}{s}-\WI{R}{e}$ coordinates in Fig.~\ref{Pix:ViewFromTop}). And finally,
type \textbf{M}: a branch with solid figures resembling a
matryoshka doll, on one side, and a quasi-spherical
body surrounded by a torus, on the other side. It must
be said that initially, we considered these two types of
configurations to be different, and only a huge number
of calculations, especially for $\sigma=2$, made it possible
to find transitional forms and combine them into one
branch.

An important finding is the conclusion that both
weak differential rotation (with $\sigma\leq 1.7$) and strongly
differential ($\sigma\geq 2.2$) do not lead to ``interesting'' configurations:
here, all solutions lie on the branch connecting
spheres with tori. All the ``exotics'' (configurations
of types \textbf{P} and \textbf{M}) lie in the vicinity of $\sigma\sim 2$.

Finally, we discuss what still needs to be done in
this area. To do this, we again refer to Fig.~\ref{Pix:MAIN2}. The first
question concerns configurations of type \textbf{M} that all
have large values of parameter $\tau$, whether they are
fragile to fragmentation and if so, what type of fragmentation
they undergo. We recall that after the loss of
axial symmetry, the development of different types of
instability is possible, in particular the so-called bar
mode or ring mode \cite{Machida2008}. In the first case, the rotating
body is stretched along an axis perpendicular to the
angular velocity vector and takes the form of a bar
(cigar) rotating almost as a solid. In the second case, a
ring is formed (for type \textbf{M}, it already exists), which
then breaks up into several separate pieces.

Besides, the it is noteworthy configurations of
types \textbf{T} and \textbf{P} lying near the critical value of $\tau\simeq 0.27$,
are they sustainable? It should be noted that the exact
criterion for fragmentation is currently unknown. In
the literature, various recipes for determining the stability
of a rotating self-gravitating configuration have
been proposed (see, for example, \cite{Christodoulou1995,Hachisu1987}). Obviously,
the answers to these questions must be given through
appropriate 3D modeling of the fragmentation process,
which the authors are currently working on.

Separately, it is necessary to mention the need for
theoretical understanding of the obtained results. It is
necessary to answer not only numerically, but also
analytically, a whole series of questions that arise when
analyzing the obtained data. Here is a list that is far
from complete: first, what exactly causes the difference
between configurations of types \textbf{T} and \textbf{P}, especially
in areas where their parameters are close?

Second, how to describe configurations of type \textbf{M}?
This apparently requires a generalization of the
Ostriker’s study \cite{Ostriker1964}, which applies only to thin tori
with $\WI{R}{e}\gg 1$ that surrounds the central object (i.e., to
configurations of type $\mathbf{M}_6$ in Fig.~\ref{Pix:Matryoshka}).

Thirdly, in Fig. \ref{Pix:ViewFromTop}, it is clear that the branches of
type \textbf{T} corresponding to different $\sigma$, all end in the area
with $\WI{M}{s}\simeq 18$ and $\WI{R}{e}\simeq 9$. What determines these particular
values? In fact, this is a question about the
structure of ``limit'' tori with $\theta\sim 0$ (see configuration $\mathbf{T}_3$
in Fig.~\ref{Pix:Tor}).

The main problem of the analytical approach is
that the application of standard methods dating back
to Chandrasekhar \cite{Chandra1933,Chandra1962} based on expansions in
terms of a small parameter (as a rule, this is the ratio of
the characteristic hydrodynamic time of the problem
and the rotation period, i.e., in order of magnitude $\omega$)
is impossible here: the rotation is strong, and the deviation
of the configuration shape from spherical is very
noticeable. In this regard, methods based on the use of
so-called oblate spheroidal coordinates \cite{Cunningham1977,Komarov1976} appear
promising; their parameters are selected so that the
density (enthalpy) isosurfaces are as close as possible
to the coordinate isolines. Work within the framework
of this approach is currently being carried out by the
authors.

\section*{ACKNOWLEDGMENTS}
The authors are grateful to the anonymous reviewer for
valuable comments that greatly contributed to improving
this article.

\section*{FUNDING}
The work of A.V. Yudin was supported by the Russian
Science Foundation, grant no. 22-12-00103.


\section*{CONFLICT OF INTEREST}

The authors of this work declare that they have no conflicts
of interest.

\clearpage







%






\end{document}